# The blood currency of suicidal mass shooters: 60 years of U.S. evidence

Quan-Hoang Vuong [1], Minh-Hoang Nguyen [1,2], Ruining Jin [3], Tam-Tri Le [4,*]

[1] Centre for Interdisciplinary Social Research, Phenikaa University, Hanoi 100803, Vietnam

[2] A.I. for Social Data Lab (AISDL), Vuong & Associates, Hanoi 100000, Vietnam

[3] Civil, Commercial and Economic Law School, China University of Political Science and Law, Beijing 100088, China

* Corresponding: tri.letam@phenikaa-uni.edu.vn

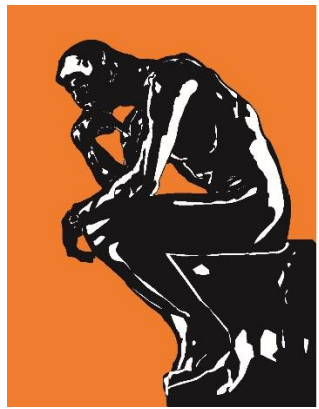

June 13, 2023

Preprint v.1.1

### Abstract

When looking at mass shooting incidents, suicidal shooters seem to carry an even more extreme sense of terror and brutality. The current study aimed to examine how mass shooters' suicidality and suicide behavioral threshold influence the severity of the mass shooting. We employed Bayesian Mindsponge Framework (BMF) analytics on a dataset of 194 mass shooters (incidents with four or more victims killed) from 1966 to 2023 in the United States (U.S.). The data were retrieved from The Violence Project Database, originally supported by the National Institute of Justice, U.S. Department of Justice. Based on the statistical analysis, we discovered that mass shooters with suicidal ideation were more likely to kill two more victims on average than their non-suicidal counterparts. For suicidal mass shooters found dead on the scene (either by self-killing or "suicide by cop"), their victim count rises by around four on average when compared to non-suicidal mass shooters. The findings were reasoned through the information-processing perspective of the Mindsponge Theory. Based on the findings and reasoning, we suggest that mass shootings should be considered within larger socio-cultural settings instead of attributing it to be driven primarily by diagnosable psychopathology. Also, promoting an appropriate interpretation of the values of life and death can be an effective way to alleviate the effects of suicidality on mass shooting severity.

**Keywords**: mass shooting; suicidality; United States; extreme violence; information processing; Bayesian Mindsponge Framework

## 1. Introduction

A mass shooting is a horrifying notion in modern society. These bloody incidents shocked the whole nation and the world each time they happened. The psychology of mass shooters is diverse, complex, and highly context-dependent, with a lot of questions still left unanswered by psychologists despite the great efforts by researchers over the years (Peterson & Densley, 2021). Alarmingly, mass shooting incidents have not only increased in recent times, but the deadliness of the attacks has also been on the rise (Densley & Peterson, 2019). Thinking about extreme violence, many people often deem suicidal attacks to be the most brutal and terror-inducing form of them all. This is intuitive and reasonable when we consider the radicalized subjective cost-benefit judgments involving the perceived values related to the act of homicide-suicide (Vuong et al., 2021). There is an exchange happening at the extreme end of the value spectrum – life and death. In the minds of suicidal mass shooters pondering upon their fatal trades, the lives they would take are the prices they desire – a payment in "blood currency".

Mass shootings, also known as active shooter incidents, have various definitions based on different sources. In the past, it lacked a clear definition from a legal perspective (Bagalman et al., 2013), but subsequent to the infamous Sandy Hook Elementary School shooting incident, The Investigative Assistance for Violent Crimes Act of 2012 defined mass killing (not mentioning the use of a firearm) as "3 or more killings" during an incident, not including the death of the perpetrator (2013). From a law enforcement perspective, the Federal Bureau of Investigation (FBI) stated that an active shooter incident refers to "one or more individuals actively engaged in killing or attempting to kill people in a populated area" through the use of a firearm (FBI, 2022). In academia, researchers define mass shooting incidents as "incidents in which four or more people are killed with a firearm in a public place within 24 hours" (Fox & Levin, 2022; Katsiyannis et al., 2018). Although definitions of mass shooting diverge, the convergence is that they all suggest mass shooting is a serious threat to society that endangers modern civilization from multiple aspects. The U.S. Congressional Research Service defines a mass public shooting as "a multiple homicide incident in which four or more victims are murdered with firearms—not including the offender(s)—within one event, and at least some of the murders occurred in a public location or locations in close geographical proximity (e.g., a workplace, school, restaurant, or other public settings), and the murders are not attributable to any other underlying criminal activity or commonplace circumstance (armed robbery, criminal competition, insurance fraud, argument, or romantic triangle)", which is also the criteria applied to the data used in this study (Peterson & Densley, 2023).

In addition to life-taking and physical harm to victims, mass shooting threats is also reflected by the significant impairments on people's mental health as well as economic conditions. Mass shooting survivors, witnesses, and their families tend to suffer from severe mental and psychological trauma, such as post-traumatic stress disorder (PTSD), major depression, anxiety, and substance abuse (Abrams, 2022; Lowe & Galea, 2017). In addition, to the general public, mass shootings also lead to fears and a declined perception of public safety in society (Lowe & Galea, 2017), as extant literature shows that mass shootings increase depression and other mental health disorders among adolescents and adults in society (Abrams, 2022; Brenan, 2019; Weitzel et al., 2023). On top of the psychological impact, mass shootings also have far-reaching economic consequences. For example, a study found that mass shootings would reduce the number of jobs and total earnings in targeted counties by about 2%, and housing prices would also decrease by 2.5% in the years following a mass shooting (Brodeur & Yousaf, 2019).

The United States (U.S.) is a country that has been plagued by mass shootings. According to the U.S. National Institute of Justice report, there is an upward trend of mass shooting numbers and effects in the past half a century in the United States, where it was found that: 1) of all the mass shooting cases, 20% took place in the last five years; 2) over a half occurred after the 00s; 3) compared with the annual casualty of 9 deaths in the 1970s, mass shootings claimed an average of 51 deaths between 2010 and 2019 (NIJ, 2022). One of the worst mass shooting incidents in the United States history is the music festival mass shooting in Las Vegas, where the perpetrator murdered 58 people and injured over 500 (NBC, 2018). Another nefarious mass shooting is the Sandy Hook Elementary School mass shooting, in which 26 people died from the shooting, including 20 children aged 6-7 and 6 school staff members (Malloy, 2015). As of June 12, 2023, the most recent mass shooting was on May 7 in Allen, Texas, where the perpetrator killed eight people and wounded seven in an outdoor shopping center (AP, 2023). The *World Population Review* listed the United States as the country with the highest number of mass shootings in the world (WorldPopulationReview, 2023). Frequent and persistent mass shootings have made a public concern regarding the solution to mass shootings and gun violence.

However, to find feasible and effective solutions, the psychology of shooters must be ascertained. According to an FBI report, because motives behind active shooters are complex and case-by-case, previous efforts to profile mass shooting perpetrators were largely unsuccessful (FBI, 2014b). The FBI also highlighted that most active shooters (80%) had a perceived grievance (not necessarily authentic), and therefore were motivated by something they saw or experienced as an injustice/unfair treatment. To this end, they wanted to conduct revenge for perceived mistreatment or unfairness. Although no single accurate profile of active shooters can be generated, there were common observable characteristics and patterns, including having a mental illness history, feeling hopelessness, seeking vengeance, and longing for notoriety (Ferguson et al., 2011; Peterson et al., 2010). In U.S. communities as well as the academic landscape, there is still a certain oversimplified view on a link between mental illnesses and mass shootings (Metzl et al., 2021). Complex interactions between individual traits (especially abnormal ones) and social factors need to be examined more systematically when investigating sources of violent criminal behavior (Peterson et al., 2022; Peterson et al., 2014).

Following these views on observable psychological characteristics, the mindset of shooters with suicidality can be examined further based on their life experiences and corresponding mental responses prior to the shootings (Rouchy et al., 2020). In an FBI study based on 160 active shooter incidents statistics, it was reported that not all active shooters had suicidality – 40% of the shooters committed suicide (FBI, 2014a). Among those suicidal shooters, the shooters who were responsible for the Columbine High School mass shooting and committed suicide after the shooting should receive greater attention (Larkin, 2009). Specifically, Larkin argued that when shooters left behind a manifesto that outlined their grievances with society, the Columbine mass shooting became a form of political protest that inspired a new wave of school shootings that were motivated by a desire to make follow-up political statements. It also gave inspiration to subsequent rampage shooters who sought to exceed the carnage numbers. The shooters gained fame and status among the outcast student subcultures. A study on homicide–suicide cases that include two or more homicide victims in China suggests that the psychological processes of the killers are influenced by local socioeconomic and cultural factors (Densley et al., 2017).

To investigate the minds of suicidal mass shooters, it is necessary to have the reasoning foundation of suicidality (including suicidal ideation and its progression into the behavioral threshold of suicide attempts). Suicide is when someone intentionally engages in self-destructive behavior with the goal of ending their life (Shneidman, 1977; WHO, 2019). Based on the impact of social connection and regulation, Emile Durkheim divided suicide into four categories: egoistic, altruistic, anomic, and fatalistic (Durkheim, 1952). According to the three-step theory (3ST) put forth by Klonsky and May (Klonsky & May, 2015), the progression from suicidal ideation to suicide attempts occurs in three stages and involves the need to commit suicide owing to pain and hopelessness, the protective power of social connections, and the ability to do so. According to the Interpersonal Theory of Suicide (ITS) (Joiner Jr et al., 2009; Van Orden et al., 2010), the combination of thwarted connectedness and perceived burdensomeness is a major contributor to suicidal thoughts. The Integrated Motivational-Volitional Model (IMV) divides the process of suicide into three phases: pre-motivational, motivational, and volitional, under the influences of biosocial, motivational, and volitional factors, respectively (O'Connor & Kirtley, 2018). The theory contends that defeat and entrapment are what trigger the emergence of suicidal ideation. According to

Rudd's Fluid Vulnerability Theory (FVT), suicide is not temporally linear but rather dynamic, and thus, suicidal ideation may manifest in episodic patterns (Rudd, 2006).

More recently, suicide is examined using a novel perspective of information processing (Nguyen et al., 2021). With the same approach, conceptual investigations into suicide attacks also expand the notion of subjective cost-benefit evaluation on the perceived value of killing oneself in the process of killing others (Vuong et al., 2021). While existing studies have not delved deep into this information processing aspect, they have provided some hints pointing to the related underlying psychological pathways. Research on mass shooters in the U.S. suggests that suicidal and non-suicidal shooters may have different personal perceptions of the value of their own lives and their future after the shootings (Lankford, 2015). U.S. perpetrators were found to have ideological and fame-seeking motives (Peterson et al., 2023; Silva, 2022). In the social contexts of developed countries, mass shootings are often carried out by lone-wolf without direct support from terrorist organizations, indicating a high degree of personal value consideration. Mass shooters driven by motivations of "personal benefits" also tend to choose places with more vulnerable people to incur more fatalities (Silva & Greene-Colozzi, 2021). In this study, using the information processing approach, we aim to explore how shooters' suicidality influence their desire to cause more human life casualty.

## 2. Methodology

### 2.1. Theoretical foundation

In this study, suicidality and the desire to cause human life casualty are examined in terms of information processing. For this purpose, we employ the mindpsonge theory – originally conceptualized to explain the process of information filtering in a new environment (Vuong & Napier, 2015) and later developed into a more general theory of dynamic information processing in the human mind (Vuong, 2023). According to the mindsponge theory, the mind is an information collection-cum-processor, influenced by multiple layers of complexity of biological and social properties and processes. The mindset is the collection of the system's accepted information, stored in memory (in the form of neural engrams). The filtering system determines what information goes in and out of the mindset, using prior accepted values as references. Both the mindset and the filtering system are altered by the act of information filtering. The trust mechanism (selective prioritization of information reception channels) can be employed to accelerate the filtering process. The fundamental biological principle that enables the dynamic adaptation of the mind differently in each person is neuroplasticity (Costandi, 2016; Eagleman, 2015). Thus, under the pressure of objective and simulated feedback from interactions with the infosphere and internal thinking, the mind optimizes its subjective values, sometimes susceptible to natural deviation in the direction of stupidity or delusion (Nguyen et al., 2023). When deviation reaches a certain threshold, the thoughts and behaviors of a person can become extremely different from social standards, but still subjectively and relatively "rational" in that individual's mind.

Suicide can be viewed as an information process involving subjective cost-benefit judgments about the duality of the perceived values of life and death (Nguyen et al., 2021). The notion of suicide, either directly (self-killing) or indirectly (actively accepting fatal impacts), negates all normal self-preservation-based motivations. However, it is still a rational mental process (from the standpoint of information processing, not social norms) employing the considerations of existing beliefs, objective observations, simulated interactions and scenarios, emotions, physiological influences, etc (Vuong et al., 2023). In normal cases of suicidal ideation, the expected benefits of ending one's life (e.g. cessation of suffering, etc.) are weighting against the negative aspects of death (e.g. discontinuation of experience, etc.) (Nguyen et al., 2021). Suicide attacks follow the same pattern of subjective cost-benefit evaluation, where one's own death is "traded" with the individual desire and satisfaction of harming the enemy (Vuong et al., 2021). The extremely violent nature of suicide bombing is built upon this mental process of fatal exchange (similar to the notion of "blood debt").

Following this line of reasoning, mass shooters who are suicidal are expected to have a higher motivation to cause as many life casualties in exchange for their own intentional death. While there is no equal exchange when it comes to human life, in the mind of the suicidal mass shooter, "sacrificing" one's own life is already the ultimate payment. Suicidality is on an intensity gradient of perceived value. Suicidal thoughts only trigger suicide attempts when this value passes an individual-specific threshold (Nguyen et al., 2021; Vuong et al., 2021). When a suicidal mass shooter committed to actually dying at the incident (not stopping nor attempting to escape), then we can assume that this behavioral threshold of suicidality has been reached, and the expected trade-off is likely more intense.

## 2.2. Materials

The data used in this study were collected, curated, and updated by The Violence Prevention Project Research Center (or in short, The Violence Project), a nonpartisan 501(c)(3) nonprofit (Peterson & Densley, 2023). The Violence Project Database was originally supported by the National Institute of Justice, U.S. Department of Justice (Award No. 2018-75-CX-0023), and has been led by two chief investigators Jillian Peterson and James Densley.

The data used in our analysis is from Version 7 (updated June 1st, 2023), including 194 mass shooters from 1966 to 2023 in the U.S. The database was constructed using public records and open-source data, including the perpetrators' diaries, "manifestos," suicide notes, social media and blog posts, audio and video recordings, interview transcripts, and personal correspondence, media coverage, documentary films and podcasts, biographies, monographs and academic journal articles, court transcripts, federal, state, and local law enforcement records, medical records, school records, and autopsy reports.

Due to the sensitive nature of the data on mass shootings, the data, codebook, and related materials are not made public, per the terms of use stated by The Violence Project. However, considering the importance of transparency and the cost of scientific research (Vuong, 2018, 2020), researchers should conduct independent cross-checking and replication studies. For academic purposes, scholars can directly request the dataset from The Violence Project at the official website (https://www.theviolenceproject.org/).

Based on the reasoning presented above, three variables are selected for analysis (see Table 1).

**Table 1**. Variable description

| Variable | Meaning | Type of variable | Value |
|---|---|---|---|
| *NumberKilled* | The number of victims killed by the shooter | Numeric | Ranging from 4 to 60 |
| *Suicidality* | Whether the shooter was suicidal | Binary | 1 is "yes"; 0 is "no" |
| *DeathOnScene* | Whether the shooter died on-scene | Binary | 1 is "yes";0 is "no" |

The variable *NumberKilled* represents the number of human victims killed by the shooter in the incident, ranging from 4 (minimum threshold according to the Congressional Research Service definition of a mass public shooting) to the highest of 60. NumberKilled has a mean value of 7.2 and a standard deviation of 6.6. The variable *Suicidality* represents the suicidality of the shooter, including whether the suicidal ideation/intention happens before or in the shooting. The variable *DeathOnScene* represents whether the shooter died in the shooting incident, including self-killing and being killed by law enforcement (or in two cases, by armed civilians).

### 2.3. Model construction

Based on the theoretical foundation presented above, we constructed the analytical model as follows. Note that the variable *DeathOnScene* is not examined on its own but conditionally in its interaction with the variable *Suicidality*. Here we test the effect of suicidality on the number of victims killed, where the intensity of such suicidal ideation reaching the behavioral threshold of actual death can be a potential moderator on the above relationship.

$$NumberKilled \sim normal(\mu, \sigma) \qquad (1)$$

$$\mu_i = \beta_0 + \beta_{Suicidality} * Suicidality_i + \beta_{Suicidality*DeathOnScene} * Suicidality_i * DeathOnScene_i \qquad (2)$$

$$\beta \sim normal(M, S) \qquad (3)$$

The probability around $\mu$ is in the form of normal distribution, where its width is the standard deviation $\sigma$. The number of victims killed by shooter $i$ is indicated by $\mu_i$. $Suicidality_i$ is the suicidality status of shooter $i$, and $DeathOnScene_i$ is the death or alive status of shooter $i$ in the shooting incident. The model has an intercept $\beta_0$ and coefficients $\beta_{Suicidality}$ and $\beta_{Suicidality*DeathOnScene}$.

A parsimonious analytical model has relatively higher predictive accuracy. To investigate a relationship using the approach of information processing, it is advantageous to construct a parsimonious model due to the high conceptual and technical compatibility with BMF analytics (Vuong, La, et al., 2022). Figure 1 shows the visualization of the model's logical network.

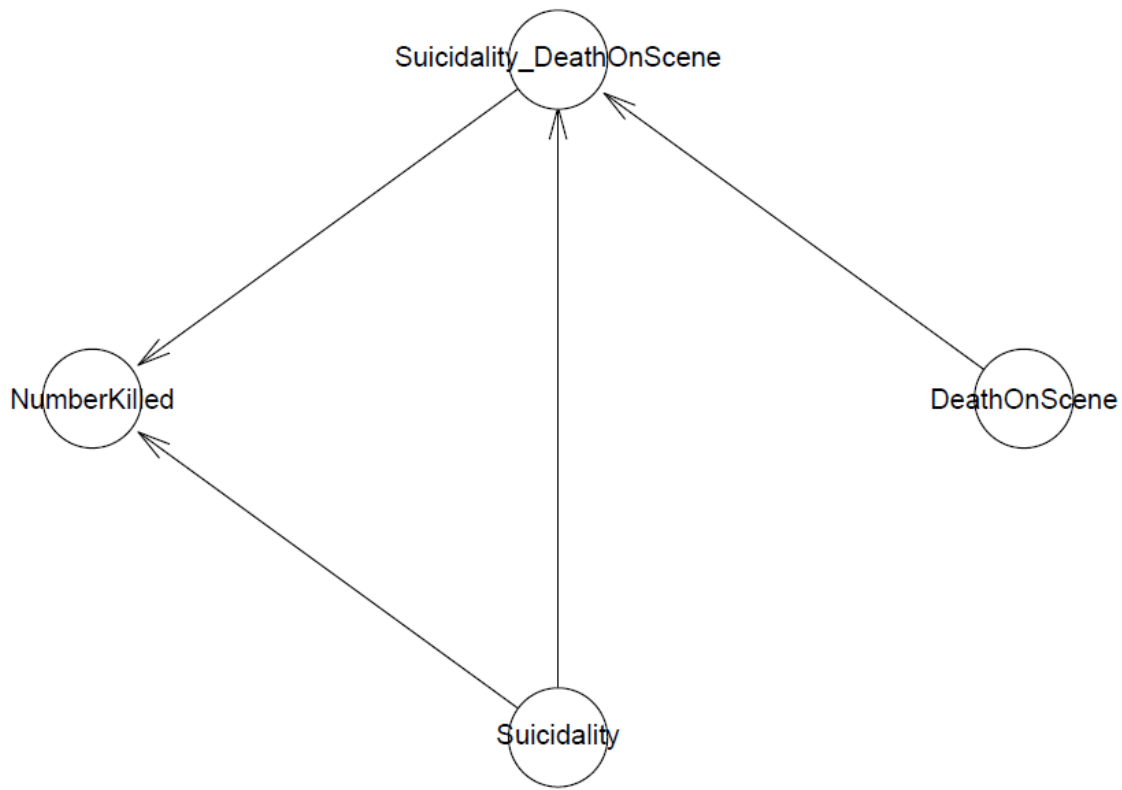

**Figure 1.** The model's logical network

### 2.4. Analysis procedure

Following the protocol of BMF analytics, our study employs Bayesian analysis assisted by the Markov Chain Monte Carlo (MCMC) technique (Nguyen et al., 2022; Vuong, La, et al., 2022). The mindsponge mechanism and Bayesian inference are highly compatible in terms of both philosophy and technicality. Bayesian inference examines all properties probabilistically, including unknown parameters. By utilizing the advantages of the MCMC algorithms

(Nguyen & Vuong, 2007; Nguyen et al., 2005), the Bayesian approach has high computational power and flexibility. MCMC-aided Bayesian analysis works particularly well with small samples (Vuong, La, et al., 2022), which is especially important in the case of this study, where due to the nature of mass shootings, the sample size is relatively small (below 200 data points) which negatively affects prediction accuracy if using the frequentist approach.

To ensure that for the analytic model, the simulated data fit well with actual data, the model's goodness-of-fit is evaluated using Pareto-smoothed importance sampling leave-one-out (PSIS-LOO) diagnostics (Vehtari et al., 2017), computed as follows.

$$LOO = -2LPPD_{loo} = -2\sum_{i=1}^{n} \log \int p(y_i|\theta)p_{post(-i)}(\theta)d\theta$$

$p_{post(-i)}(\theta)$ is the posterior distribution based on the data minus data point $i$. In the "LOO" package in R, $k$-Pareto values are used in the PSIS method for computing leave-one-out cross-validation. Observations with $k$-Pareto values greater than 0.7 are considered to have a high degree of influence on the PSIS estimate. A model is commonly considered fit when its $k$ values are below 0.5. Furthermore, we continue to check the model's goodness-of-fit using Graphical posterior predictive checks (PPCs). The posterior predictive distribution formula is as follows.

$$p(y^{rep}|y) = \int p(y^{rep}|\theta)p(\theta|y)d\theta$$

$y$ is the observed data, $\theta$ is the vector of parameters, $y^{rep}$ is the simulated data. With this, we can compare the simulated batches of parameters to their reference distribution. Next, we can measure the discrepancy between the model and the data based on experimental quantities that evaluate aspects of the data using test quantities. In MCMC simulation, we compare test quantities on observed data after each step $T(y, \theta(s))$ with predictive test quantities $T(y^{rep(s)}, \theta(s))$ respectively. If the discrepancy is negligible, we can say that the model captures the actual data well. The posterior predictive $p$-value is calculated in the simulation steps as follows with $n_{sims}$ as the number of iterations and $\theta$ is a parameter's vector.

$$p = \frac{1}{n_{sims}} \sum_{s=1}^{n_{sims}} 1(T(y^{rep(s)}, \theta(s)) > T(y, \theta(s)))$$

The effective sample size ($n\_eff$) and the Gelman-Rubin shrink factor ($Rhat$) are used to statistically analyze convergence. The $n\_eff$ value represents the number of iterative samples that are not autocorrelated during stochastic simulation. If $n\_eff$ is more than 1000, the effective samples are adequate for accurate inference. The convergence of Markov chains can also be evaluated using the $Rhat$ value (Gelman shrink factor). If the value is more than 1.1, the model likely does not converge. If $Rhat$ equals 1, the model can be deemed well-convergent. Additionally, trace plots, Gelman-Rubin-Brooks plots, and autocorrelation plots are used to visually check the convergence.

We employ the **bayesvl** R package (La & Vuong, 2019) to conduct Bayesian analysis. The package is open-source, has high visualization capabilities, and has straightforward operation (Vuong, Nguyen, et al., 2022). The MCMC setup for the model consists of 5000 iterations, including 2000 warmup iterations and four chains. Besides using uninformative prior to avoid subjective influence on the results, we utilize the prior-tweaking technique to test the robustness of the model. If the posterior results show consistent patterns across different priors (uninformative, prior disbelief in the effect of suicidality, and prior belief in the positive influence of suicidality on the victim number), then the model can be deemed robust.

### 3. Results

The latest model fitting run was conducted on June 11, 2023, with a total elapsed time of 57.4 seconds (uninformative), 44.1 seconds (disbelief in the effect), and 55.6 seconds (belief in the effect), on R version 4.2.1, and Windows 11. Table 2 shows the estimated results using uninformative prior.

The PSIS diagnostic plot (see Figure 2) shows that most computed *k*-values are below 0.5. However, there are a few observations with *k*-values over 0.7.

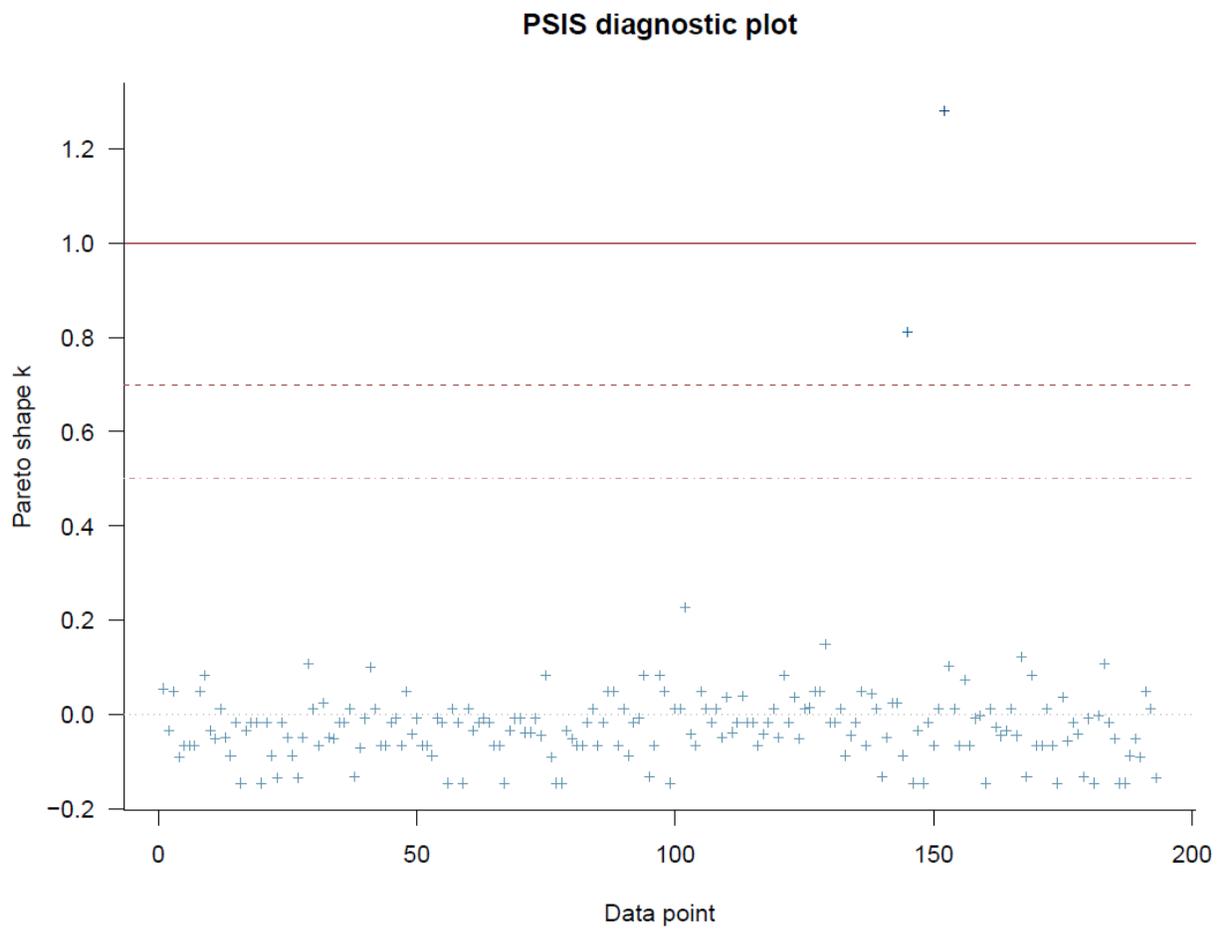

**Figure 2.** The model's PSIS-LOO diagnostic plot

To further check the distributions of simulated data computed from the posterior predictive distribution. Figure 3 shows how 50 simulated datasets fit actual data.

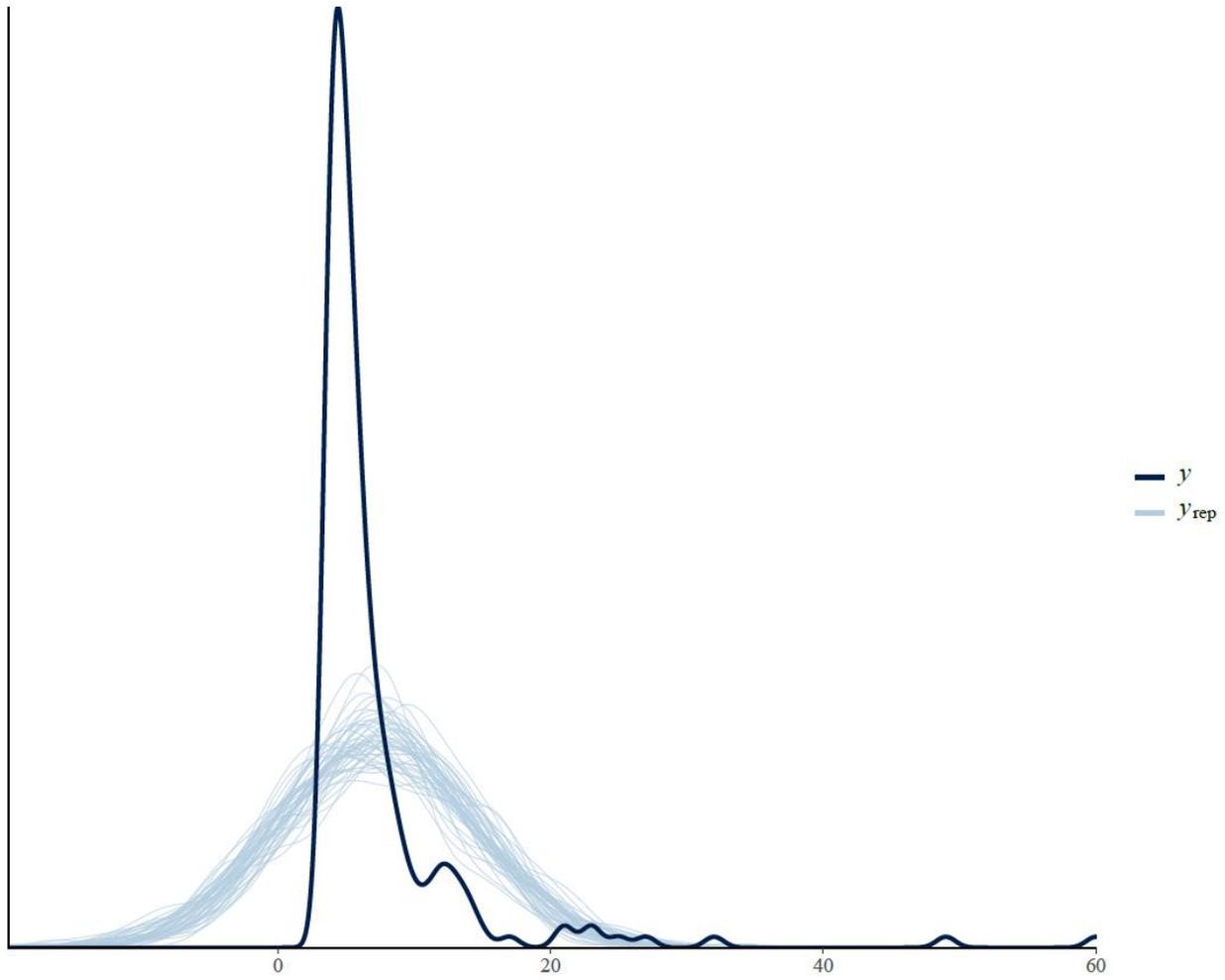

**Figure 3.** Graphical posterior predictive checks of the model

We can also see the distribution of the mean (log) of NumberKilled of simulated datasets compared to the mean in the actual data (see Figure 4). Both Figure 3 and 4 indicate that the model has an acceptable goodness-of-fit.

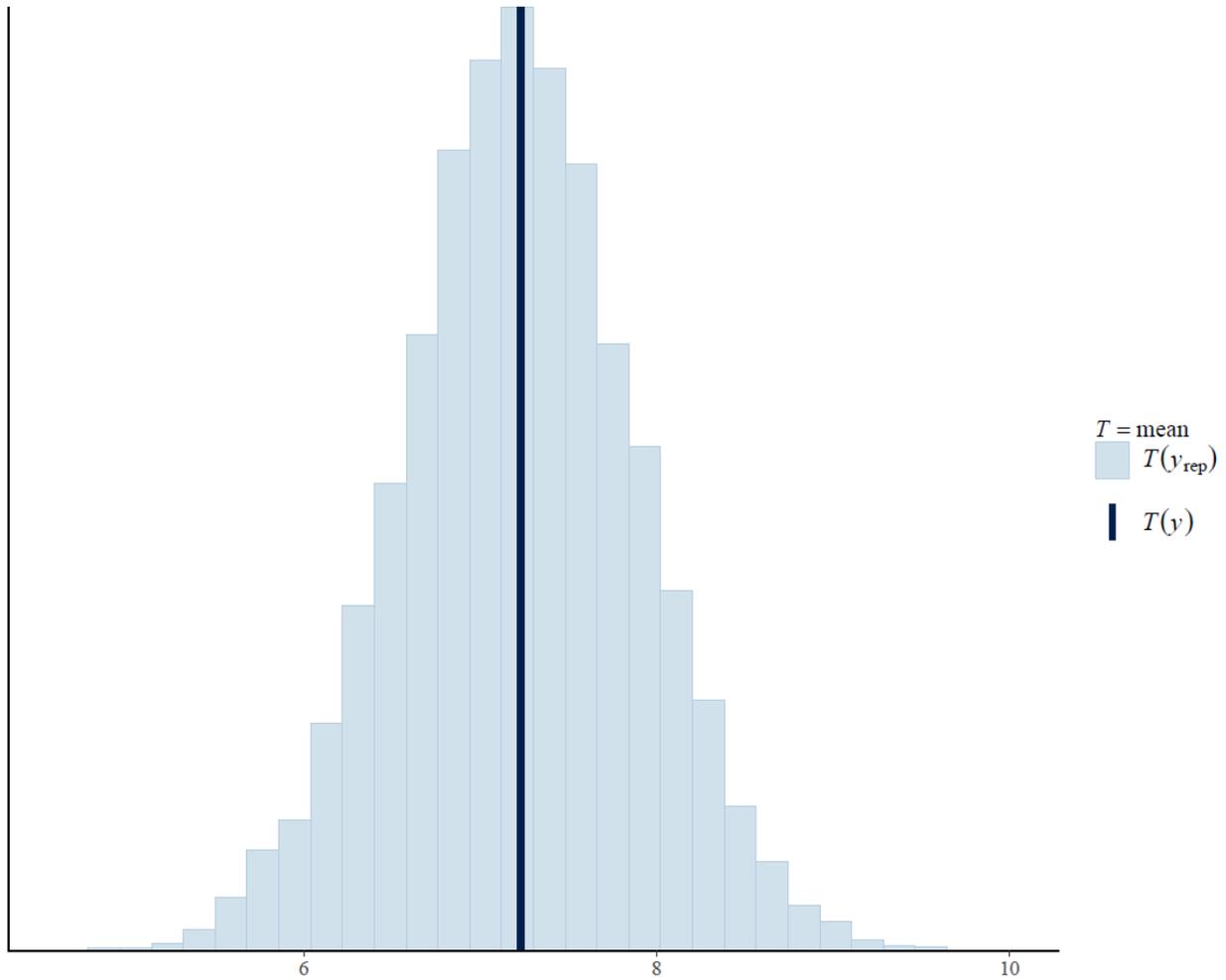

**Figure 4.** Visualized test quantities of the model

Table 2 shows the statistical results of the analyses using uninformative and informative priors. The *n_eff* and *Rhat* values for all parameters are healthy (n_eff > 1000 and Rhat = 1), indicating the convergence of the Markov chains. The results using uninformative prior, prior reflecting disbelief in the effect, and prior reflecting the belief in the effect all show consistent patterns of effects for the coefficients, meaning that the model is robust.

**Table 2:** Simulated posteriors of the analytical model

| Parameters | Uninformative prior | | | | Informative prior | | | | | | | |
| --- | --- | --- | --- | --- | --- | --- | --- | --- | --- | --- | --- | --- |
| | | | | | Disbelief in the effect | | | | Belief in the effect | | | |
| | Mean | SD | n_eff | Rhat | Mean | SD | n_eff | Rhat | Mean | SD | n_eff | Rhat |
| *Constant* | 5.14 | 0.83 | 8066 | 1 | 5.35 | 0.80 | 7944 | 1 | 4.97 | 0.79 | 8080 | 1 |
| *Suicidality* | 1.76 | 1.28 | 7009 | 1 | 1.28 | 1.11 | 7530 | 1 | 2.14 | 1.08 | 7095 | 1 |
| *Suicidality\*DeathOnScene* | 1.78 | 1.19 | 7497 | 1 | 2.04 | 1.13 | 8247 | 1 | 1.57 | 1.10 | 8238 | 1 |

The trace plots (see Figure 5) show that the Markov chains converge well, fluctuating around equilibriums after the warm-up period (from the 2000th iteration onward). The Gelman-Rubin-Brooks plots (see Figure 6) show that the shrink factors drop to 1 during the warmup period. In the autocorrelation plots (see Figure 7), the autocorrelation levels among iterative samples are eliminated as the values drop to zero after some finite lags.

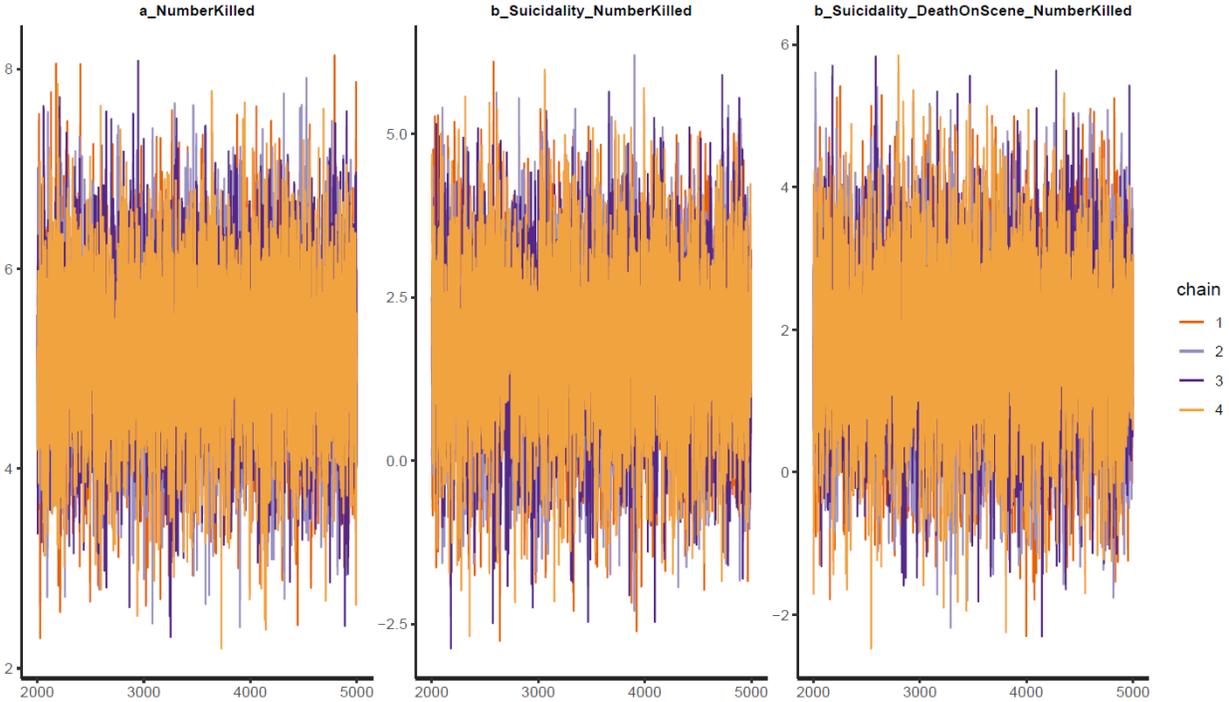

**Figure 5.** Trace plots for the analytical model

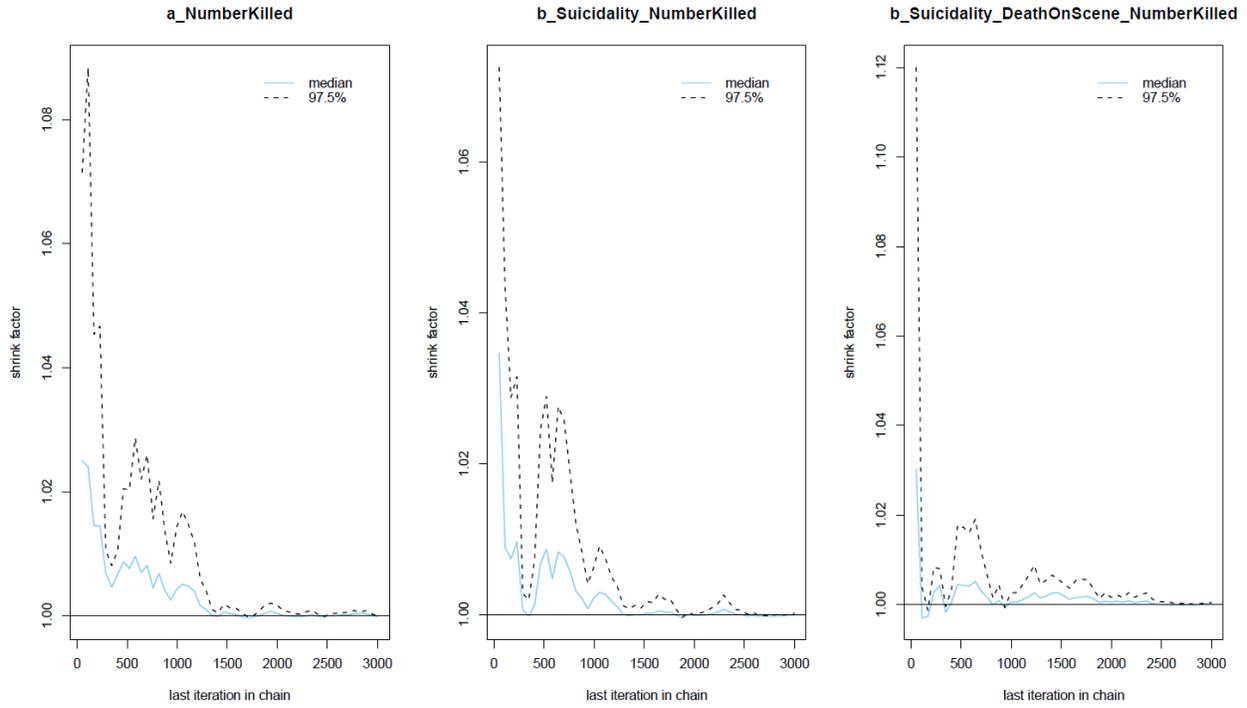

**Figure 6.** Gelman-Rubin-Brooks plots for the analytical model

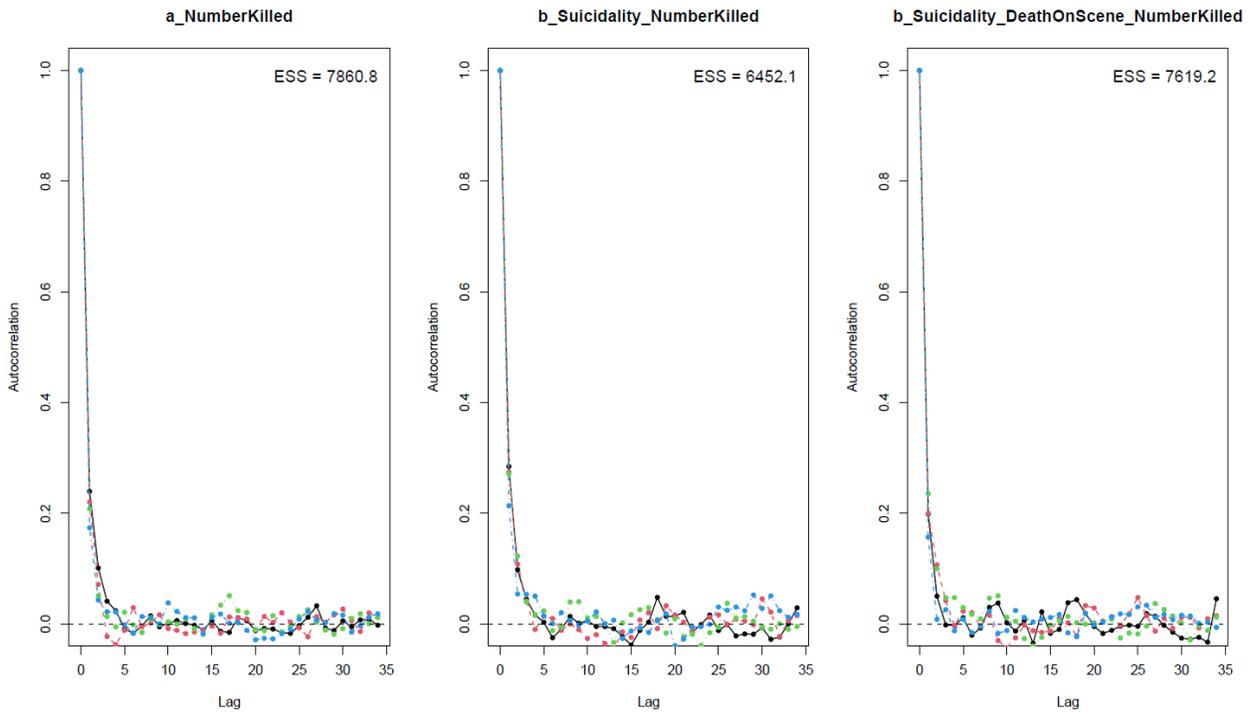

**Figure 7.** Autocorrelation plots for the analytical model

As shown in Table 2, *Suicidality* is positively associated with *NumberKilled* (for the analysis using uninformative prior, $M_{Suicidality} = 1.76$ and $SD_{Suicidality} = 1.28$). *Suicidality\*DeathOnScene* is also positively associated with *NumberKilled* (for the analysis using uninformative prior, $M_{Suicidality*DeathOnScene} = 1.78$ and $SD_{Suicidality*DeathOnScene} = 1.19$). This means that *DeathOnScene* has a positive moderating effect on the association between *Suicidality* and *NumberKilled*. The visualization of the posterior distributions is presented in Figure 8, where the thick blue line represents the credible range within the 89% Highest Posterior Density Intervals. Both the effects of *Suicidality* and *Suicidality\*DeathOnScene* are reliable, as most of their distributions stay on the positive side.

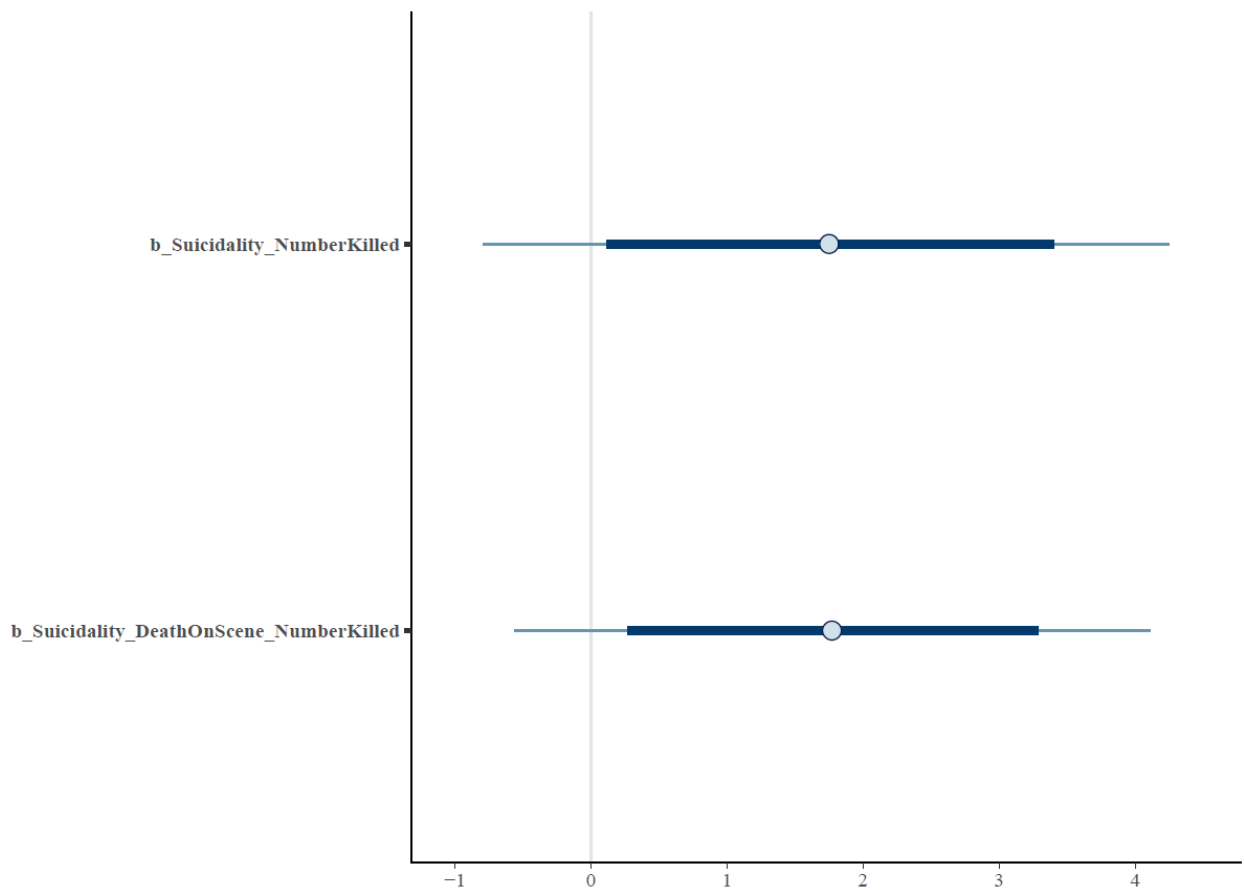

**Figure 8.** Posteriors' distributions on an interval plot

To aid result interpretation, Figure 7 illustrates the estimated outcomes based on posterior results. Compared to a shooter without suicidal ideation, a shooter with suicidality reaching the suicide behavioral threshold (dying on-scene) would kill approximately four more people (see the blue line in Figure 7).

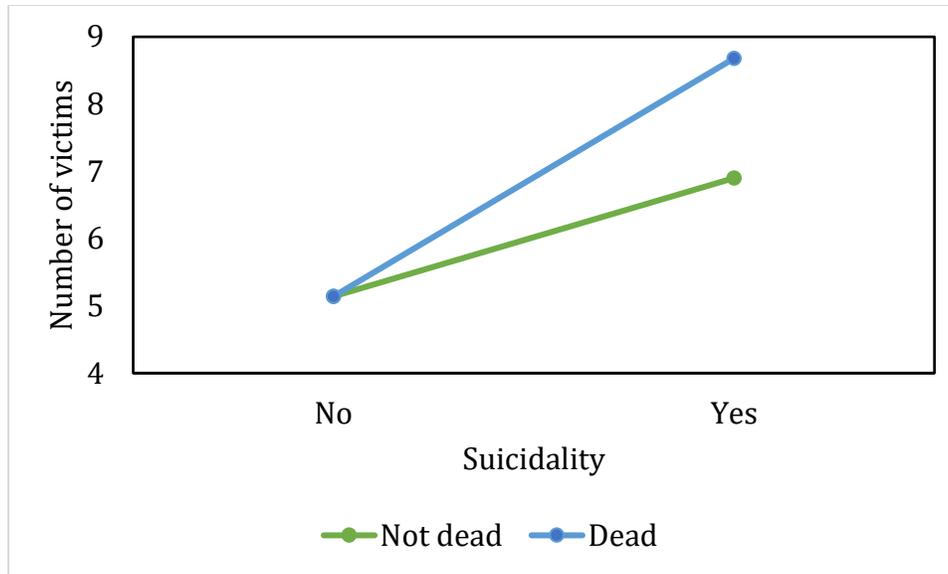

**Figure 9.** Estimated number of victims based on the shooter's suicidality and suicide behavioral threshold

## 4. Discussion

The current study explored whether the mass shooters' suicidal ideation and death-on-scene outcome are associated with the number of victims. We employed the Mindsponge Theory to provide a theoretical foundation for the study and conducted BMF analytics on the dataset retrieved from The Violence Project Database to validate our reasoning. The dataset included the records of 194 mass shooters from 1996 to 2023 in the U.S. Based on the statistical analysis, we found that if the mass shooters have suicidal ideation, they are more likely to kill two more victims on average. If a suicidal mass shooter is found dead on-scene (suicidal ideation reaching behavioral threshold), the number of victims is found to increase by around four on average, compared to non-suicidal mass shooters.

These results support our theoretical reasoning that the number of victims of mass shooters is influenced by the mass shooters' subjective cost-benefit judgments. In the mind of the mass shooters, killing people is deemed beneficial. This evaluation is driven by their value systems of which the priorities are shaped by trusted values (or beliefs). Such trusted values can include ideological beliefs, grievances and revenge fantasies, incels worldview, and desire for infamy (Anisin, 2022; Cottee, 2021; Knoll, 2010; Silva & Greene-Colozzi, 2019). With one or some of these trusted values in mind, the act of killing can be considered an act of fulfilling their goals and the number of killed victims can be considered a measure of achievement, or a type of "blood currency". In other words, the higher number of victims they kill, the higher satisfaction/fulfillment due to the accomplishment they might expect to feel. Moreover, among all subjective value perceptions, one's own life is normally considered the most prioritized core mental construct in most people. A suicidal mass shooter would want to have a "worthy trade" of this ultimate self-destructive energy in terms of perceived value: the lives of other people.

According to the Mindsponge Theory, when the act of killing is considered a beneficial action and the outcome can be measured by the "blood currency", the mass shooters will try to maximize the acquired "currency." Logically, the mass shooters' "achievements" are positively associated with how well they invest in planning and conducting the mass shooting event. The investment can include the availability of weapons, selection of shooting location and condition, motivations, etc. (Lankford, 2015; Silva & Greene-Colozzi, 2021). For mass shooters with suicidal ideation, they might consider their life as one of the resource types that they can invest in conducting the mass shooting while planning and executing to optimize the received outcome (i.e., being armed with more weapons and attacking at different locations) (Lankford, 2015). This can help explain how mass shooters with suicidal ideation are more likely to kill more victims.

However, ideation is not necessarily translated to actual behavior, because there might be instinctual desires for survival or other different thoughts and beliefs acting as behavioral inhibitors (through diminishing the perceived value of the act of suicide). For mass shooters with the actual behavior of suicide (either by self-killing or "suicide by cop"), they need a strong determination and motivation that can sufficiently suppress and surpass other instinctual or rational desires to live. In other words, they become even more radicalized than shooters with suicidal ideation but still struggle to decide whether to suicide or not. In such circumstances, other types of information in the shooters' minds will be considered irrelevant and excluded, while information supporting the optimization of the killing action is further reinforced, increasing the effectiveness and efficiency of killing plans and preparation (i.e., preparing more weapons effective at killing, finding more suitable locations, etc.). Following this reasoning, we can explain how suicidal mass shooters with on-scene death are found to kill more victims than those with suicidal mass shooters but are alive after the incidents.

From this study's findings, it is suggested that mass shooters' suicidal ideation and behaviors are positively associated with the severity of mass shooting events in terms of human casualties. However, such found effects of suicidal ideation and behavior should be viewed as assisting factors intensifying the severity of the mass shooting rather than the primary driver of mass shootings. From the perspective of mindsponge thinking, mass shooters' mindsets need to contain information that establishes value systems that grant the act of killings merits, such as for fulfilling ideological beliefs, revenge fantasies, desire of infamy, etc. Holding suicidal desires only lessens the value of being alive, subsequently making the mass-shooter-to-be more likely to accept the costs of mass shootings and more committed to planning and executing the shooting. Thus, we advocate the strategy of abandoning the assumption that acts of mass violence are driven solely by diagnosable psychopathology, and instead consider such destructive motivations within larger social structures and cultural settings (Metzl et al., 2021; Peterson & Densley, 2017; Peterson et al., 2014; Skeem et al., 2011).

To alleviate the effects of suicidality on mass shooting severity, it is imperative to reduce the number of people with suicidal ideation among the normal population (considering the unexpectedness of mass shooting incident emergence). Promoting the values of life can be an effective way to increase the safeguard against suicidal thoughts. It should be noted that prohibiting access to death-related information might not be an effective safeguarding method because of two reasons. Firstly, death is an indispensable part of the human life cycle, so sooner or later a person will have to process death-related values in one way or another. Secondly, a strict prohibition approach can temporarily prevent people from viewing death as an option in their cost-benefit evaluation in most normal contexts. However, when being exposed to death-related information, people with a limited understanding of life-and-death values might be more vulnerable to suicide. It is evident that people's religious affiliation was associated with a higher likelihood of suicidal ideation when their accessibility to help-related information from social sources was thwarted (Vuong et al., 2021). This is probably because of their exposure to life-death information in religious teaching. However, ideation does not necessarily lead to action. Although religious affiliation is found to be not protective against suicidal ideation, it is protective against suicide attempts and suicide (Lawrence et al., 2016).

This study has some limitations. Although The Violence Project has made considerable efforts in data collection and fact-checking, there are certain limitations due to data privacy laws in the U.S. Some cases in the past might not have been well-reported or influenced by biases from media agencies. Another limitation is the small sample size due to the nature of mass shootings (it is a good thing that this number is not too big). Despite this drawback, Bayesian analysis aided by the MCMC technique allows for relatively reliable prediction when working with this small sample size.

## Acknowledgment

The data used in this study was kindly provided by The Violence Project (https://www.theviolenceproject.org). We thank the team at The Violence Project, chief investigators Dr. Jillian Peterson and Dr. James Densley.